\documentclass[reprint,superscriptaddress,10pt,a4paper,preprintnumbers,nofootinbib]{revtex4-1}
\usepackage{amsmath,amssymb,latexsym}
\usepackage{bm}
\usepackage{graphicx}
\usepackage{hepunits}
\usepackage[svgnames]{xcolor}

\newcommand{\nn}{\nonumber}

\newcommand{\mK}{\mathcal{K}}

\allowdisplaybreaks

\begin{document}

\title{Alphabet of one-loop Feynman integrals}

\author{Jiaqi Chen}
\email{chenjq@ihep.ac.cn}
\affiliation{Institute of High Energy Physics, Chinese Academy of Sciences, Beijing 100049, China}
\author{Chichuan Ma}
\email{chichuanma@pku.edu.cn}
\affiliation{School of Physics and State Key Laboratory of Nuclear Physics and Technology,\\
Peking University, Beijing 100871, China}
\author{Li Lin Yang}
\email{yanglilin@zju.edu.cn}
\affiliation{Zhejiang Institute of Modern Physics, Department of Physics, Zhejiang University, Hangzhou 310027, China}

\begin{abstract}
In this paper, we present the universal structure of the alphabet of one-loop Feynman integrals. The letters in the alphabet are calculated using the Baikov representation with cuts. We consider both convergent and divergent cut integrals, and find that letters in the divergent cases can be easily obtained from the convergent cases by taking certain limits. The letters are written as simple expressions in terms of various Gram determinants. The knowledge of the alphabet makes it easy to construct the canonical differential equations of the $d\log$ form, and helps to bootstrap the symbols of the solutions.
\end{abstract}


\maketitle

\section{Introduction}

The systematic study of one-loop Feynman integrals in perturbative quantum field theories dates back to the end of the 1970s when 't~Hooft and Veltman \cite{tHooft:1978jhc} calculated the generic one-, two-, three- and four-point scalar integrals in dimensional regularization (DREG) up to order $\epsilon^0$ , where $\epsilon=(4-d)/2$ with spacetime dimension $d$. Passarino and Veltman \cite{Passarino:1978jh} then demonstrated that tensor integrals up to four points can be systematically reduced to scalar ones, and later it was shown \cite{Bern:1992em, Bern:1993kr} that integrals with more than four external legs in $4-2\epsilon$ dimensions can be expressed as lower-point ones up to order $\epsilon^0$. These developments in principle solved the problem of next-to-leading order (NLO) calculations for tree-induced scattering processes. 

The improvements of experimental precision and the progress of theoretical studies require the understanding of scattering amplitudes and cross sections at higher orders in perturbation theory. In this respect, we need to compute the one-loop integrals to higher orders in $\epsilon$. These allow us to predict the infrared divergences appearing in the two-loop amplitudes \cite{Catani:1998bh, Aybat:2006wq, Aybat:2006mz, Sterman:2002qn, Becher:2009cu, Becher:2009qa, Becher:2009kw, Ferroglia:2009ep, Ferroglia:2009ii}, and are also necessary for computing the one-loop squared amplitudes which are essential ingredients of next-to-next-to-leading order (NNLO) cross sections.

Unlike the terms up to order $\epsilon^0$, generic results for the higher order terms are not available yet. Part of the reason is that integrals with more than four external legs are in general not reducible to lower-point ones when considering higher orders in $\epsilon$. These require further calculations which are often rather complicated due to the increasing number of physical scales involved.

It is known \cite{Bourjaily:2019exo, Herrmann:2019upk, Chen:2020uyk} that one-loop integrals in a given family admit a uniform transcendentality (UT) basis satisfying canonical differential equations of the form \cite{Henn:2013pwa}
\begin{equation}
d\vec{f}(\vec{x},\epsilon) = \epsilon \, d\bm{A}(\vec{x}) \, \vec{f}(\vec{x},\epsilon) \, ,
\label{eq:can_de}
\end{equation}
where $\vec{x}$ is the set of independent kinematic variables, and the matrix $d\bm{A}$ takes the $d\log$-form:
\begin{equation}
d\bm{A}(\vec{x}) = \sum_i \bm{C}_i \, d\log(W_i(\vec{x})) \, .
\end{equation}
In the above expression, $\bm{C}_i$ are matrices consisting of rational numbers, and $W_i(\vec{x})$ are algebraic functions of the variables. The functions $W_i$ are called the ``letters'' for this integral family, and the set of all independent letters is called the ``alphabet''.

At one-loop, a canonical basis can be generically constructed by looking for $d\log$-form integrands \cite{ArkaniHamed:2010gh, Gehrmann:2011xn, Drummond:2013nda, Henn:2013pwa, Arkani-Hamed:2014via, Bern:2014kca, Herrmann:2019upk, Bourjaily:2019exo, Chen:2020uyk, Chen:2022lzr}. On the other hand, obtaining the $d\log$ matrix $d\bm{A}(\vec{x})$ is not always a trivial task when the number of variables is large. We note that the $d\log$ matrix can be easily reconstructed if we have the knowledge of the alphabet $\{W_i(\vec{x})\}$ in advance, since the coefficient matrices $\bm{C}_i$ can then be obtained by bootstrapping.

Having the alphabet (and hence the matrix $d\bm{A}(\vec{x})$) in a good form also helps to solve the differential equations \eqref{eq:can_de} order-by-order in the dimensional regulator $\epsilon$. The (suitably normalized) solution can be written as a Taylor series:
\begin{equation}
\vec{f}(\vec{x},\epsilon) = \sum_{n=0}^{\infty} \epsilon^n \, \vec{f}^{(n)}(\vec{x}) \, ,
\end{equation}
where the $n$th order coefficient function can be written as a Chen iterated integral \cite{Chen:1977oja}
\begin{align}
\vec{f}^{(n)}(\vec{x}) =& \int_{\vec{x}_0}^{\vec{x}} d\bm{A}(\vec{x}_n) \cdots \int_{\vec{x}_0}^{\vec{x}_2} d\bm{A}(\vec{x}_1)
 + \vec{f}^{(n)}(\vec{x}_0) \, .
\end{align}
Such iterated integrals can be analyzed using the language of ``symbols'' \cite{Brown:2009qja, Goncharov:2010jf, Duhr:2011zq} that encodes the algebraic properties of the resulting functions. In certain cases, these iterated integrals can be solved analytically (either by direct integration or by bootstrapping). The results can often be written in terms of generalized polylogarithms (GPLs) \cite{Goncharov:1998kja} which allow efficient numeric evaluation~\cite{Vollinga:2004sn, Naterop:2019xaf, Wang:2021imw}. When an analytic solution is not available, it is straightforward to evaluate them numerically either by numerical integration or by a series expansion~\cite{Moriello:2019yhu, Hidding:2020ytt}.

In this paper, we describe a generic method to construct the letters systematically from cut integrals in the Baikov representation \cite{Baikov:1996iu, Frellesvig:2017aai}. The letters can be generically written in terms of various Gram determinants. The letters and symbols of one-loop integrals have been considered in \cite{Arkani-Hamed:2017ahv, Abreu:2017ptx, Abreu:2017enx, Abreu:2017mtm}, and our method is similar to that in \cite{Abreu:2017ptx, Abreu:2017enx, Abreu:2017mtm}. Nevertheless, we evaluate the cut integrals differently and obtain equivalent but simpler expressions in certain cases utilizing properties of Gram determinants. Furthermore, we consider the cases of divergent cut integrals which were ignored in earlier studies. Using our results, it is easy to write down all letters for a given integral family even before constructing the differential equations. These letters will also appear in the corresponding two-loop integrals.

\section{The canonical basis of one-loop integrals}

We use the method of \cite{Chen:2020uyk, Chen:2022lzr} to construct the canonical basis in the Baikov representation. In this section, we give a brief review of the construction procedure since it will also be relevant for obtaining the alphabet in the matrices $d\bm{A}(\vec{x})$.

Consider a generic one-loop integral topology with $N = E + 1$ external legs, where $E$ is the number of independent external momenta. Integrals in this topology can be written as
\begin{align}
\label{eq:intI}
I_{a_1, \cdots, a_N} = \int \frac{d^dl}{i \pi^{d/2}} \frac{1}{z_1^{a_1} z_2^{a_2}\cdots z_N^{a_N}} \,,
\end{align}
where $z_i$ are the propagator denominators given by
\begin{gather}
\label{eq:propagator_general}
z_1=l^2-m_1^2 \,, \quad z_2= (l+p_1)^2-m_2^2 \,, \quad \cdots \,, \nn \\
\quad z_N= (l+p_1+\cdots+p_{E})^2-m_N^2 \,.
\end{gather}
Here $p_1,\ldots,p_E$ are external momenta which we assume to span a space-like subspace of the $d$-dimensional Minkowski spacetime. This corresponds to the so-called (unphysical) Euclidean kinematics. Results in the physical phase-space region can be defined by analytic continuation.

The idea of the Baikov representation is to change the integration variables from loop momenta $l^\mu$ to the Baikov variables $z_i$, and the result is given by
\begin{multline}
I_{a_1,\ldots,a_N} = \frac{1}{(4 \pi)^{E/2} \, \Gamma\big((d-E)/2\big)} \\
\times \int_{\mathcal{C}} \frac{ \left| G_N(\bm{z}) \right|^{(d-E-2)/2}}{\left| \mathcal{K}_N \right|^{(d-E-1)/2}  }  \prod_{i=1}^{N} \frac{dz_i}{z_i^{a_i}} \, ,
\label{eq:1loopbaikov}
\end{multline}
where $\bm{z}=\{z_1,\ldots,z_N\}$ is the collection of the Baikov variables. The function $G_N(\bm{z})$ is a polynomial of the $N$ variables, while $\mathcal{K}_N$ is independent of $\bm{z}$. They are given by
\begin{equation}
G_N(\bm{z}) \equiv G(l,p_1,\ldots,p_E) \, , \quad \mathcal{K}_N = G(p_1, \cdots, p_{E}) \, ,
\label{eq:GN}
\end{equation}
where the Gram determinant is defined as
\begin{equation}
\label{eq:gram1}
G(q_1,\ldots,q_n) \equiv  \det
\begin{pmatrix}
q_1 \cdot q_1 & q_1 \cdot q_2 & \cdots & q_1 \cdot q_n
\\
q_2 \cdot q_1 & q_2 \cdot q_2 & & \vdots
\\
\vdots & & \ddots & \vdots
\\
q_n \cdot q_1 & \cdots & \cdots & q_n \cdot q_n
\end{pmatrix}
\, .
\end{equation}
Note that in Eq.~\eqref{eq:GN}, the scalar products involving the loop momentum $l$ should be re-expressed in terms of $\bm{z}$:
\begin{align}
l^2 &= z_1 + m_0^2 \, , \nonumber
\\
l \cdot p_i &= \frac{z_{i+1} + m_{i+1}^2 - p_i^2 - z_i - m_i^2}{2} - \sum_{j=1}^{i-1} p_i \cdot p_j \, .
\end{align}
The integration domain $\mathcal{C}$ in Eq.~\eqref{eq:1loopbaikov} is determined by the condition $G_N(\bm{z})/\mathcal{K}_N \leq 0$ with Euclidean kinematics.

We are now ready to write down the UT integrals $g_N$ for any $N$ according to \cite{Chen:2020uyk}. We need to distinguish between the cases of odd $N$ and even $N$:
\begin{align}
g_{N} \big|_{\text{$N$ odd}} &= \frac{\epsilon^{(N+1)/2}}{(4\pi)^{(N-1)/2} \, \Gamma(1-\epsilon)} \nn \\
&\times \int \left( -\frac{\mK_N}{G_N(\bm{z})} \right)^\epsilon \prod_{i=1}^N \frac{dz_i}{z_i} \, , \nn
\\
g_{N} \big|_{\text{$N$ even}} &= \frac{\epsilon^{N/2}}{(4\pi)^{(N-1)/2} \, \Gamma(1/2-\epsilon)} \nn \\
&\times \int \frac{\sqrt{G_N(\bm{0})}}{\sqrt{G_N(\bm{z})}} \left( -\frac{\mK_N}{G_N(\bm{z})} \right)^\epsilon \prod_{i=1}^N \frac{dz_i}{z_i} \, ,
\label{eq:UTint}
\end{align}
where we set $\mK_1 = 1$, and $\bm{0}$ means that all $z_i$'s are zero.
Note that $g_{2n-1}$ and $g_{2n}$ can be naturally identified as Feynman integrals in $2n-2\epsilon$ dimensions:
\begin{align}
g_N \big|_{N=2n-1} &= \epsilon^n \sqrt{\mK_N} \, I_{1 \times N}^{(2n-2\epsilon)} \,, \nn
\\
g_N \big|_{N=2n} &= \epsilon^n \sqrt{G_N(\bm{0})} \, I_{1 \times N}^{(2n-2\epsilon)} \,,
\label{eq:UTMIs}
\end{align}
where $I^{(d)}_{1 \times N}$ denotes the $d$-dimensional $N$-point Feynman integral with all powers $a_i = 1$:
\begin{align}
I^{(d)}_{1 \times N} \equiv \int \frac{d^dl}{i \pi^{d/2}} \frac{1}{z_1 z_2 \cdots z_N} \,.
\end{align}
They can be related to Feynman integrals in $4-2\epsilon$ dimensions using the dimensional recurrence relations \cite{Tarasov:1996br, Lee:2009dh}. Applying the above to all sectors of a family, we build a complete canonical basis satisfying $\epsilon$-form differential equations.

\section{Letters in differential equations: convergent cases}
\label{sec:letter}

Given a basis of Feynman integrals, it is straightforward to calculate the derivatives with respect to some kinematic variable $x_i$. For a UT basis $\vec{f}(\vec{x},\epsilon)$ we write
\begin{equation}
\frac{\partial}{\partial x_i} \vec{f}(\vec{x},\epsilon) = \epsilon \, \bm{A}_i(\vec{x}) \, \vec{f}(\vec{x},\epsilon) \, ,
\end{equation}
where the elements in the matrix $\bm{A}_i(\vec{x})$ have the property that they only contain simple poles. In principle, one may already attempt to solve these differential equations by direct integration. However, this is usually rather difficult when $\bm{A}_i(\vec{x})$ contains many irrational functions (square roots). Therefore it is often very useful to combine the partial derivatives into a total derivative, and rewrite the differential equations in the form of Eq.~\eqref{eq:can_de}. To do that it is important to know the alphabet (i.e., the set of independent letters $W_i(\vec{x})$) in the matrix $d\bm{A}(\vec{x})$. With the knowledge of the alphabet, it is straightforward to reconstruct the whole matrix $d\bm{A}(\vec{x})$ by comparing the coefficients in the partial derivatives.

In principle, one may obtain the letters by directly integrating the matrices $\bm{A}_i(\vec{x})$ over the variables $x_i$, and manipulating the resulting expressions. However, in the presence of many square roots (containing high-degree polynomials) in multi-scale problems, these integrations are not easy to perform, and the results are often extremely complicated. One may find examples in various one-loop and multi-loop calculations, e.g., Refs.~\cite{Heller:2019gkq, Bonciani:2019jyb, Chen:2022nxt}.
With such kind of expressions, it is highly non-trivial to decide whether a set of letters are independent or not.\footnote{There is a package \texttt{SymBuild} \cite{Mitev:2018kie} which can carry out such a task, but the computational burden is rather heavy when there are many square roots.} Furthermore, from experience we know that letters involving square roots can often be written in the form
\begin{equation}
\frac{P(\vec{x}) - \sqrt{Q(\vec{x})}}{P(\vec{x}) + \sqrt{Q(\vec{x})}} \, ,
\end{equation}
where $P$ and $Q$ are polynomials. Such letters have nice properties under analytic continuation: they are real when $Q(\vec{x}) > 0$, and become pure phases when $Q(\vec{x}) < 0$. But it is not easy to recover this structure from direct integration.

Given the above considerations, we now describe a novel method to obtain the letters, especially those with square roots and multiple scales. Our method is based on the $d\log$-form integrals in the Baikov representation under various cuts. We will work with the generic propagator denominators in Eq.~\eqref{eq:propagator_general} and the Baikov representation \eqref{eq:1loopbaikov}. Without loss of generality, we define the Baikov cut on the first $r$ variable $z_1,\ldots,z_r$ by \cite{Frellesvig:2017aai}
\begin{multline}
\label{eq:baikov_cut}
I_{a_1,\ldots,a_N}\big|_{\text{$r$-cut}} = \frac{1}{(4 \pi)^{E/2} \, \Gamma((d-E)/2)}  \\
\times \int  \prod_{j=r+1}^N \frac{dz_j}{z_j^{a_j}} \prod_{i=1}^r \oint_{z_i=0} \frac{dz_i}{z_i^{a_i}} \frac{| G_N(\bm{z}) |^{(d-E-2)/2}}{|\mathcal{K}_N|^{(d-E-1)/2}  }   \, .
\end{multline}
An important property of the Baikov cut is that if one of the powers $a_i$ $(1 \leq i \leq r)$ is non-positive, the cut integral vanishes according to the residue theorem. The coefficient matrices in the differential equations are invariant under the cuts, and we will utilize this fact to obtain the letters by imposing various cuts.

To begin with, we write the differential equation satisfied by an $N$-point one-loop UT integral $g_N$ (see Eqs.~\eqref{eq:UTint} and \eqref{eq:UTMIs}) as
\begin{align}
\label{eq:DE-letter}
dg_N(\vec{x},\epsilon) &= \epsilon \, dM_N(\vec{x}) \, g_N(\vec{x},\epsilon) \nn \\
&+ \epsilon \sum_{m < N} \sum_i  dM_{N,m}^{(i)}(\vec{x}) \,   g_m^{(i)}(\vec{x},\epsilon) \, ,
\end{align}
where $g_N(\vec{x},\epsilon)$ and $g_m^{(i)}(\vec{x},\epsilon)$ are components of the canonical basis $\vec{f}(\vec{x},\epsilon)$, while $dM_N(\vec{x})$ and $dM_{N,m}^{(i)}(\vec{x})$ are entries in the matrix $d\bm{A}(\vec{x})$. It should be clear from the above equation that the derivative of $g_N$ cannot depend on higher-point integrals, and cannot depend on other $N$-point integrals as well. It may depend on several $m$-point integrals for each $m < N$, and we use a superscript like in $g_m^{(i)}$ and $dM_{N,m}^{(i)}$ to distinguish them. These $m$-point integrals can be obtained by ``squeezing'' some of the propagators in the $N$-point diagram. 

From Eq.~\eqref{eq:DE-letter}, one sees that it's possible to focus on a particular entry of the $d\bm{A}$ matrix by imposing some cuts. We elaborate on this in the following. In this Section we will assume that the master integrals (after imposing cuts) have no divergences, such that the integrands can be expanded as Taylor series in $\epsilon$ before integration. It can be shown that in this situation only $g_N$, $g_{N-1}^{(i)}$ and $g_{N-2}^{(i)}$ appear in the right side of Eq.~\eqref{eq:DE-letter}. It turns out that the most complicated letters are given by these cases. Occasionally we encounter divergences in the cut integrals, and one has to expand the integrands as Laurent series in terms of distributions. We will discuss these cases in the next Section.

\subsection{The self-dependence $dM_{N}$}
\label{sec:self}

The self-dependent term in Eq.~\eqref{eq:DE-letter} is easy to extract by imposing the ``maximal-cut'', i.e., cut on all variables $\bm{z}$. All the lower-point integrals vanish under this cut, and the differential equation becomes
\begin{equation}
d\tilde{g}_{N}(\vec{x},\epsilon) = \epsilon \, dM_N(\vec{x}) \, \tilde{g}_{N}(\vec{x},\epsilon) \, ,
\end{equation}
where $\tilde{g}_N$ denotes the cut integral. Using the generic form of UT integrals in Eq.~\eqref{eq:UTint}, it is easy to see that
\begin{equation}
dM_N(\vec{x}) = d\log \left(-\frac{\mK_N(\vec{x})}{\widetilde{G}_N(\vec{x})}\right) ,
\label{eq:dM_N}
\end{equation}
where
\begin{equation}
\widetilde{G}_N(\vec{x}) \equiv G_N(\bm{0}) \, .
\end{equation}
Hence we see that the corresponding letter can be chosen as
\begin{equation}
W_N(\vec{x}) = \frac{\widetilde{G}_N(\vec{x})}{\mK_N(\vec{x})} \, .
\end{equation}
We note that two letters are equivalent if they only differ by a constant factor or a constant power, i.e.,
\begin{equation}
W(\vec{x}) \sim c \, W(\vec{x}) \sim \left[ W(\vec{x}) \right]^n \, .
\label{eq:letter_equiv}
\end{equation}
Therefore in practice, we may choose a form that is convenient for the particular case at hand.

It is possible that $G_N(\bm{0}) = 0$ such that $W_N(\vec{x}) = 0$ and cannot be a letter. In this case, the integral $\tilde{g}_{N}$ itself vanishes under the maximal cut. This means that the integral is reducible to integrals in sub-sectors, and we don't need to consider it as a master integral.

\subsection{Dependence on sub-sectors with one fewer propagator}

We now consider the dependence of the derivative of $g_N$ on sub-sectors with $N-1$ propagators. There can be $N$ such sub-sectors, corresponding to ``squeezing'' one of the $N$ propagators. Focusing on one of the sub-sector integral $g_{N-1}^{(i)}$, we can always reorganize the propagators (by shifting the loop momentum and relabel the external momenta) such that the squeezed one is $z_N$. We can then impose cut on the first $N-1$ variables, and write the differential equation as
\begin{align}
d\tilde{g}_{N}(\vec{x},\epsilon) &= \epsilon \, dM_{N}(\vec{x}) \, \tilde{g}_{N}(\vec{x},\epsilon) \nn
\\
&+ \epsilon \, dM_{N,N-1}(\vec{x}) \, \tilde{g}_{N-1}(\vec{x},\epsilon) \,,
\end{align}
where we have suppressed the superscript since only one sub-sector survives the cut. The letter in $dM_{N}(\vec{x})$ has been obtained in the previous step, and we now need to calculate the letter in $dM_{N,N-1}(\vec{x})$.

\subsubsection{Odd number of propagators}

We first consider the case where $N$ is an odd number. Using the generic form of one-loop UT integrals Eq.~\eqref{eq:UTint}, we can write
\begin{multline}
d\int_{r_-}^{r_+} \left(-\frac{\mK_N}{G_N(\bm{0}',z_N)}\right)^\epsilon \frac{dz_N}{z_N} \\
= \epsilon \, dM_N \int_{r_-}^{r_+} \left(-\frac{\mK_N}{G_N(\bm{0}',z_N)}\right)^\epsilon \frac{dz_N}{z_N}
\\
+ dM_{N,N-1} \, \frac{2^{1-2\epsilon} \, \Gamma^2(1-\epsilon)}{ \Gamma(1-2\epsilon)} \left(-\frac{\mK_{N-1}}{\widetilde{G}_{N-1}}\right)^\epsilon \, ,
\end{multline}
where the integration boundary is determined by the two roots $r_{\pm}$ of the polynomial $G_N(\bm{0}',z_N)$, and $\bm{0}'$ means that the vector $\bm{z}'\equiv\{z_1,\ldots,z_{N-1}\}$ is zero.

If both $r_+$ and $r_-$ are non-zero, the integration over $z_N$ is convergent for $\epsilon \to 0$. We can then set $\epsilon = 0$ in the equation and get
\begin{equation}
dM_{N,N-1} = \frac{1}{2} \, d\int_{r_-}^{r_+} \frac{dz_N}{z_N} = \frac{1}{2} \, d\log\frac{r_+}{r_-} \, .
\end{equation}
We may already set the letter to $r_+/r_-$ and stop at this point. However, it will be useful to write $r_\pm$ in terms of certain Gram determinants. This not only simplifies the procedure to compute the letter, but also tells us about the physics in the divergent situations $r_+ = 0$ or $r_- = 0$.

Given the propagator denominators \eqref{eq:propagator_general} and the definition of the Gram determinant \eqref{eq:gram1}, it is easy to see that $z_N$ only appears in the top-right and bottom-left corners of the Gram matrix. Using the expansion of the determinant in terms of cofactors, we can write
\begin{align}
G_N(\bm{0}',z_N)=-\frac{1}{4}\mK_{N-1} z_N^2 - \widetilde{B}_N z_N +\widetilde{G}_N \, ,
\end{align}
where $\widetilde{B}_N \equiv B_N(\bm{0})$ with (recall that $E=N-1$)
\begin{align}
B_N(\bm{z}) \equiv G(l,p_1,\ldots,p_{E-1};\, p_{E},p_1,\ldots,p_{E-1})\, ,
\end{align}
Here we have defined an extended Gram determinant
\begin{multline}
G(q_1,\ldots,q_n;\, k_1,\ldots,k_n) \\
= \det
\begin{pmatrix}
q_1 \cdot k_1 & q_1 \cdot k_2 & \cdots & q_1 \cdot k_n
\\
q_2 \cdot k_1 & q_2 \cdot k_2 & & \vdots
\\
\vdots & & \ddots & \vdots
\\
q_n \cdot k_1 & \cdots & \cdots & q_n \cdot k_n
\end{pmatrix}
.
\end{multline}

We may further use the geometric picture of Gram determinants to simplify the two roots. The Gram determinants can be expressed as
\begin{align}
G(q_1,\ldots,q_n) &= \det \left( q_i^\mu q_j^\nu g_{\mu\nu} \right) \nn
\\
&= \det(g_{\mu\nu}) \left[ V(q_1,\ldots,q_n) \right]^2 \, ,
\end{align}
where $q_i^\mu$ is the $\mu$th component of $q_i$ in the subspace spanned by $\{q_1,\ldots,q_n\}$ (with an arbitrary coordinate system), and $g_{\mu\nu}$ is the metric tensor of this subspace. $V(q_1,\ldots,q_n)$ is the volume of the parallelotope formed by the vectors $q_1,\ldots,q_n$ (in the Euclidean sense). 

Let $l^\star$ denote a solution to the equation $\bm{z} = 0$ (recall that $z_i$ contain scalar products involving the loop momentum $l$), we can write
\begin{align}
\widetilde{G}_{N-1} &= G(l^\star,p_1,\ldots,p_{E-1}) \, , \nonumber
\\
\widetilde{G}_N &= G(l^\star,p_1,\ldots,p_E) \, , \nonumber
\\
\widetilde{B}_N &= G(l^\star,p_1,\ldots,p_{E-1};\, p_{E},p_1,\ldots,p_{E-1}) \, .
\end{align}
We let $l^\star_\perp$ and $p_{E\perp}$ to denote the components of $l^\star$ and $p_E$ perpendicular to the subspace spanned by $p_1,\ldots,p_{E-1}$, respectively. We are working in the region that the subspace of external momenta is space-like, and $l^\star_\perp$ must be time-like (since $l^\star$ is either time-like or light-like due to $(l^\star)^2 - m_1^2 = 0$). We can write the components of $l^\star_\perp$ perpendicular and parallel to $p_{E\perp}$ as $|l^\star_\perp|\cosh(\eta)$ and $|l^\star_\perp|\sinh(\eta)$, respectively, where $|l^\star_\perp| \equiv \sqrt{(l^\star_\perp)^2}$. We also denote $|p_{E\perp}| \equiv \sqrt{-p_{E\perp}^2}$. These allow us to write 
\begin{align}
\frac{\widetilde{B}_N}{\mK_{N-1}} &= - |l_{\perp}^\star| |p_{E\perp}| \sinh(\eta) \,,  \quad \frac{\mK_{N}}{\mK_{N-1}} = -|p_{E\perp}|^2 \,, \nn
\\
\frac{\widetilde{G}_N}{\mK_{N-1}} &= -|l_{\perp}^\star|^2 |p_{E\perp}|^2 \cosh^2(\eta) \,, \quad \frac{\widetilde{G}_{N-1}}{\mK_{N-1}} = |l_{\perp}^\star|^2 \,.
\label{eq:triangle1}
\end{align}
It then follows that
\begin{equation}
\widetilde{B}_N^2 + \mK_{N-1} \widetilde{G}_N = -\mK_{N-1}^2 |l_{\perp}^\star|^2 |p_{E\perp}|^2 = \mK_{N} \widetilde{G}_{N-1} \, .
\end{equation}
Note that the above relation can also be obtained from the Sylvester's determinant identity applied to Gram determinants (for other applications of this relation, see, e.g., \cite{Chen:2020uyk, Dlapa:2021qsl, Chen:2022lzr}). We will encounter further instances of this relation later in this work.

Expressing $r_\pm$ in terms of the Gram determinants, we can finally write the letter in $dM_{N,N-1}$ (for odd $N$) as
\begin{equation}
W_{N,N-1}(\vec{x}) = \frac{\widetilde{B}_{N}-\sqrt{\widetilde{G}_{N-1}\mathcal{K}_N}}{\widetilde{B}_{N}+\sqrt{\widetilde{G}_{N-1}\mathcal{K}_N}} \, .
\label{eq:W_n1_odd}
\end{equation}
We emphasize that the ingredients $\widetilde{B}_N$, $\widetilde{G}_{N-1}$ and $\mK_N$ can be very complicated functions of the kinematic variables $\vec{x}$ when $N$ and the length of $\vec{x}$ are large, and it is not easy to obtain the letter through direct integration in multi-scale problems.

If one of $r_\pm$ is zero, the integration over $z_N$ is divergent when $\epsilon \to 0$, and we cannot expand the integrand as a Taylor series. Actually, one can see that $W_{N,N-1}(\vec{x})$ in Eq.~\eqref{eq:W_n1_odd} becomes zero in this situation. On the other hand, this requires $\widetilde{G}_N = 0$, which means that $g_N$ vanishes under maximal cut, and hence is not a master integral. It is also possible that $\widetilde{G}_{N-1} = 0$ and $g_{N-1}$ is not a master. In this case $\log W_{N,N-1} = \log(1) = 0$ drops out of the differential equations. We therefore do not need to consider these cases here. Similar considerations apply to the $N$-even case coming next.

\subsubsection{Even number of propagators}

We now turn to the situation where $N$ is an even number. We proceed similarly as the odd case, and arrive at the cut differential equation
\begin{align}
&d \int_{r_-}^{r_+} \frac{dz_N}{z_N} \frac{\sqrt{\widetilde{G}_N}}{\sqrt{G_N(\bm{0}',z_N)}} \left[-\frac{\mK_N}{G_N(\bm{0}',z_N)}\right]^\epsilon \nonumber
\\
&= \epsilon \, dM_{N} \int_{r_-}^{r_+} \frac{dz_N}{z_N} \frac{\sqrt{\widetilde{G}_N}}{\sqrt{G_N(\bm{0}',z_N)}} \left[-\frac{\mK_N}{G_N(\bm{0}',z_N)}\right]^\epsilon \nonumber
\\
&+ 2\pi \, \epsilon \, \frac{2^{2\epsilon} \, \Gamma(1-2\epsilon)}{\Gamma^2(1-\epsilon)} \, dM_{N,N-1} \left(-\frac{\mK_{N-1}}{\widetilde{G}_{N-1}}\right)^\epsilon \, .
\end{align}
We again assume that the integration over $z_N$ is convergent for $\epsilon \to 0$. We can then expand the integrands on both sides of the above equation. At order $\epsilon^0$, the integral on the left side is
\begin{equation}
\int_{r_-}^{r_+} \frac{dz_N}{z_N} \frac{\sqrt{\widetilde{G}_N}}{\sqrt{G_N(\bm{0}',z_N)}} = i \pi \, .
\label{eq:pi}
\end{equation}
Hence its derivative is zero. Comparing the order $\epsilon^1$ coefficients, and plugging in the form of $dM_N$ obtained earlier in Eq.~\eqref{eq:dM_N}, we get
\begin{multline}
dM_{N,N-1} = -\frac{1}{2\pi} \, d \int_{r_-}^{r_+} \frac{dz_N}{z_N} \frac{\sqrt{\widetilde{G}_N}}{\sqrt{G_N(\bm{0}',z_N)}}
\\
\times \log \frac{G_N(\bm{0}',z_N)}{\widetilde{G}_N} \, .
\end{multline}

The above integrand involves multi-valued functions such as square roots and logarithms. To define the integral, we need to choose a convention including branch cuts for these functions and also the path from $r_-$ to $r_+$. Different conventions will lead to results differing by some constants or an overall minus sign, but these do not affect the letter up to the equivalence mentioned in Eq.~\eqref{eq:letter_equiv}.

\begin{figure}[t!]
\centering
\includegraphics[width=0.4\textwidth]{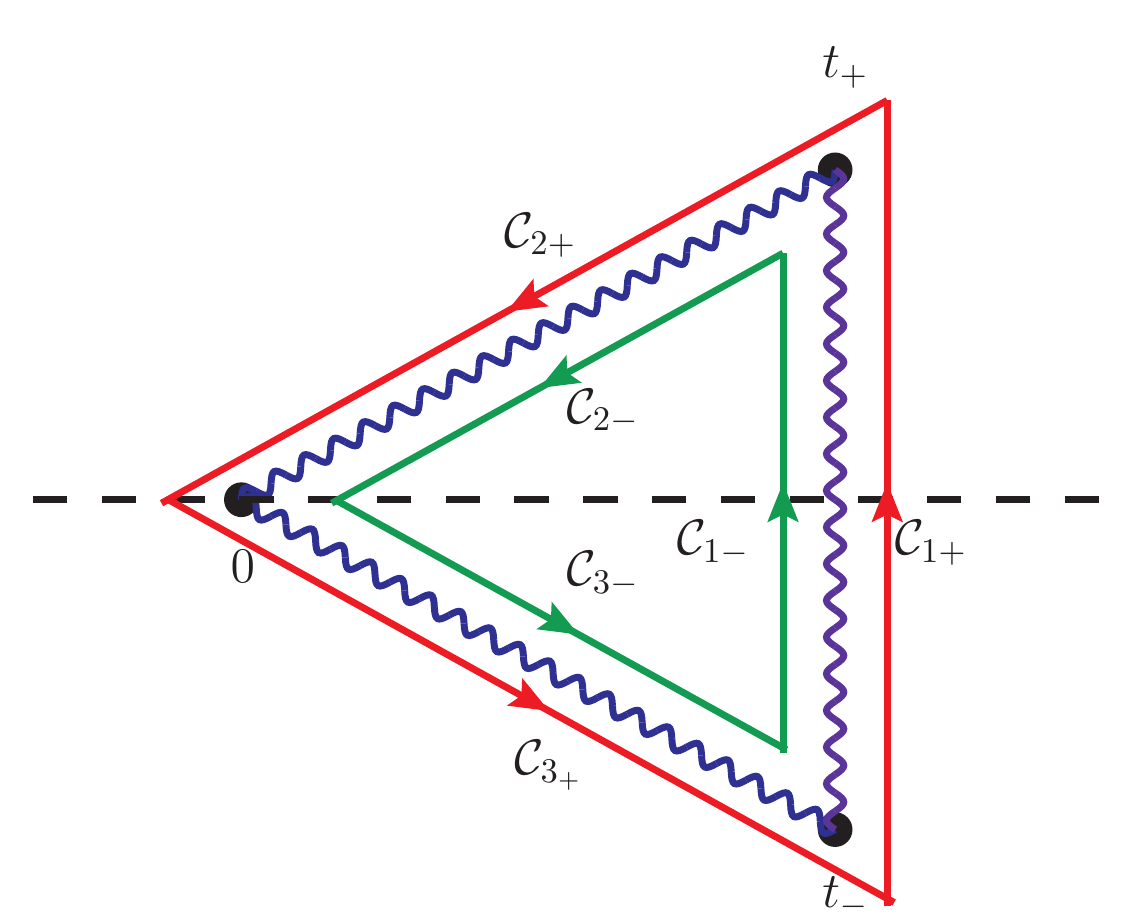}
\caption{\label{fig:contour_M43}The branch cuts and integration paths for $M_{N,N-1}$ with even $N$.}
\end{figure}

We denote $G_N(\bm{0}',z_N)$ as $(r_+-z_N)(z_N-r_-) \mK_{N-1}/4$ with $\mK_{N-1} > 0$, and write the integral as
\begin{multline}
M_{N,N-1} = -\frac{1}{2\pi} \, \int_{r_-}^{r_+} \frac{dz_N}{z_N} \sqrt{\frac{r_+ r_-}{(z_N-r_+)(z_N-r_-)}}
\\
\times \log \frac{(z_N-r_+)(z_N-r_-)}{r_+ r_-} \, .
\end{multline}
The branch cuts involve the points $r_\pm$ and $\infty$ on the complex $z_N$ plane. To represent the cuts more clearly, we perform the change of variable:
\begin{align}
z_N = \frac{1}{t} \,, \quad t_\pm = \frac{1}{r_\mp} \, .
\end{align}
The branch points then become $t_\pm$ and $0$, and we write the integral as
\begin{equation}
M_{N,N-1} = -\frac{1}{2\pi} \int_{t_-}^{t_+}  I(t) \, dt \, ,
\label{eq:M43_dt}
\end{equation}
with the integrand
\begin{align}
I(t) &= \frac{1}{\sqrt{(t-t_+)(t-t_-)}} \left[ \log\frac{t-t_+}{t} + \log\frac{t-t_-}{t} \right] .
\label{eq:M43_It}
\end{align}
With this form of the integrand, we choose the branch cut for the square root to be the line segment between $t_+$ and $t_-$, and the branch cuts for the two logarithms to be the line segments between $0$ and $t_\pm$, respectively. These branch cuts are depicted as the wiggly lines in Fig.~\ref{fig:contour_M43}, together with several paths $C_{i\pm}$ which lie infinitesimally close to the cuts. We define the square root following the convention that $\sqrt{(t-t_+)(t-t_-)} \to t$ when $t \to \infty$.

We choose the integration path in Eq.~\eqref{eq:M43_dt} to along the line segment $\mathcal{C}_{1+}$, and write the integral as
\begin{equation}
M_{N,N-1} = -\frac{1}{4\pi} \left[ \int_{\mathcal{C}_{1+}} I(t) \, dt - \int_{\mathcal{C}_{1-}} I(t) \, dt \right] ,
\end{equation}
where we have used the fact that the values of $I(t)$ on $\mathcal{C}_{1\pm}$ differ by a sign. Since there are no other singularities in the complex $t$ plane (including $\infty$), we may deform the paths as long as we don't go across the branch cuts. Hence we know that
\begin{align}
M_{N,N-1} &= \frac{1}{4\pi} \left[ \int_{\mathcal{C}_{2+}} I(t) \, dt - \int_{\mathcal{C}_{2-}} I(t) \, dt \right] \nn \\
&+ \frac{1}{4\pi} \left[ \int_{\mathcal{C}_{3+}} I(t) \, dt - \int_{\mathcal{C}_{3-}} I(t) \, dt \right].
\end{align}
On the paths $\mathcal{C}_{2+}$ and $\mathcal{C}_{2-}$, there is a $2\pi i$ difference coming from the first logarithm in Eq.~\eqref{eq:M43_It}. A similar difference of $-2\pi i$ arising from the second logarithm is there between $\mathcal{C}_{3+}$ and $\mathcal{C}_{3-}$. Therefore we have
\begin{align}
dM_{N,N-1} &= -\frac{i}{2} \, d\int_0^{t_+} \frac{dt}{\sqrt{(t-t_+)(t-t_-)}} \nn \\
&\quad - \frac{i}{2} \, d\int_0^{t_-} \frac{dt}{\sqrt{(t-t_+)(t-t_-)}} \nonumber
\\
&= -i \, d\log \frac{\sqrt{r_+} - \sqrt{r_-}}{\sqrt{r_+} + \sqrt{r_-}} \, .
\end{align}
Note that with the above convention, we have
\begin{equation}
\int_{r_-}^{r_+} \frac{dz_N}{z_N} \frac{\sqrt{\widetilde{G}_N}}{\sqrt{G_N(\bm{0}',z_N)}} = \int_{t_-}^{t_+} \frac{dt}{\sqrt{(t-t_+)(t-t_-)}} = i \pi \, .
\end{equation}

We can now express the roots $r_\pm$ in terms of Gram determinants. The result can be written as
\begin{align}
dM_{N,N-1} = \frac{i}{2} \, d\log \frac{\widetilde{B}_{N}-\sqrt{-\widetilde{G}_{N} \mK_{N-1}}}{\widetilde{B}_{N}+\sqrt{-\widetilde{G}_{N} \mK_{N-1}}} \, ,
\end{align}
where the definition of $\widetilde{B}_{N}$, $\widetilde{G}_{N}$ and $\mK_{N-1}$ is similar as before.
Hence we can write the letter in $dM_{N,N-1}$ (for even $N$) as
\begin{equation}
W_{N,N-1}(\vec{x}) = \frac{\widetilde{B}_{N}-\sqrt{-\widetilde{G}_{N} \mK_{N-1}}}{\widetilde{B}_{N}+\sqrt{-\widetilde{G}_{N} \mK_{N-1}}} \, .
\label{eq:W_n1_even}
\end{equation}
As mentioned earlier, we don't need to consider the divergent case $\widetilde{G}_{N-1}=0$ or the trivial case $\widetilde{G}_N=0$ here.

\subsection{Dependence on sub-sectors with two fewer propagators}

As in the previous subsection, we consider the dependence of the derivative of $g_N$ on sub-sectors with $N-2$ propagators. Without loss of generality, we cut on the variables $\bm{z}'=\{z_1,\ldots,z_{N-2}\}$. Now we are left with two sub-sectors with $N-1$ propagators: one with $\bm{z}',z_{N-1}$ and the other with $\bm{z}',z_{N}$. We use a superscript to distinguish these two, and the differential equation then reads
\begin{align}
d\tilde{g}_{N} &= \epsilon \Big( dM_{N} \, \tilde{g}_{N} + dM_{N,N-1}^{(1)} \, \tilde{g}_{N-1}^{(1)} \nn \\
&\quad + dM_{N,N-1}^{(2)} \, \tilde{g}_{N-1}^{(2)} + dM_{N,N-2} \, \tilde{g}_{N-2} \Big) \, ,
\end{align}
where we have suppressed the arguments of the functions for simplicity.

\subsubsection{Odd number of propagators}\label{sec:W53}

If $N$ is an odd number, assuming convergence and expanding the integrands, we find
\begin{align}
&d\int_\mathcal{C} \frac{dz_{N-1}}{z_{N-1}}\frac{dz_N}{z_N} = 4\pi \, dM_{N,N-2} \nn \\
&+ 2 \, dM_{N,N-1}^{(1)} \int_{r_{-}^{(1)}}^{r_{+}^{(1)}} \frac{dz_{N-1}}{z_{N-1}} \frac{\sqrt{\widetilde{G}_{N-1}^{(1)}}}{\sqrt{G_{N-1}^{(1)}(\bm{0}',z_{N-1})}} \nonumber\\
&+ 2 \, dM_{N,N-1}^{(2)} \int_{r_{-}^{(2)}}^{r_{+}^{(2)}} \frac{dz_N}{z_N} \frac{\sqrt{\widetilde{G}_{N-1}^{(2)}}}{\sqrt{G_{N-1}^{(2)}(\bm{0}',z_N)}} \,,
\end{align}
where the domain $\mathcal{C}$ is determined by $G_N(\bm{0}',z_{N-1},z_N)\geq 0$, and $r_{\pm}^{(i)}$ are the two roots of the polynomial $G_{N-1}^{(i)}(\bm{0}',z)$.

The two integrals on the right-hand side can be easily performed using Eq.~\eqref{eq:pi}, and we have
\begin{align}
dM_{N,N-2} = dI_{N,N-2} - \frac{i}{2} \left( dM_{N,N-1}^{(1)} + dM_{N,N-1}^{(2)} \right)\, , 
\end{align}
where $I_{N,N-2}$ is the double integral
\begin{align}
I_{N,N-2} = \frac{1}{4\pi} \int_\mathcal{C} \frac{dz_{N-1}}{z_{N-1}}\frac{dz_N}{z_N} \, . 
\end{align}
The integration domain $\mathcal{C}$ is controlled by the positivity of the polynomial
\begin{align}
G_N(\bm{0}',z_{N-1}, z_N) &= -\frac{1}{4}\mK_{N-1} z_N^2 - B_N(\bm{0}',z_{N-1},0) \, z_N \nn \\
&\quad + G_N(\bm{0}',z_{N-1},0) \, .
\end{align}
The integration over $z_N$ can be easily performed to arrive at
\begin{align}
I_{N,N-2} &= \frac{1}{4\pi} \int_{r_{N-1,-}}^{r_{N-1,+}} I(z_{N-1}) \, dz_{N-1} \nonumber
\\
&\equiv \frac{1}{4\pi} \int_{r_{N-1,-}}^{r_{N-1,+}} \frac{dz_{N-1}}{z_{N-1}} \nn \\
&\quad \times \log \frac{B_N(\bm{0}',z_{N-1},0) - \sqrt{\Delta(z_{N-1})}}{B_N(\bm{0}',z_{N-1},0) + \sqrt{\Delta(z_{N-1})}} \,, 
\label{eq:I53}
\end{align}
where $r_{N-1,\pm}$ are the two roots of the polynomial
\begin{equation}
G_{N-1}^{(1)}(\bm{z}',z_{N-1}) = G(l,p_1,\ldots,p_{E-1}) \, ,
\end{equation}
and
\begin{align}
\Delta(z_{N-1}) &= \left[ B_N(\bm{0}',z_{N-1},0) \right]^2 + \mK_{N-1} G_N(\bm{0}',z_{N-1},0) \nn \\
&= \mK_N G_{N-1}^{(1)}(\bm{0}',z_{N-1}) \,.
\end{align}

\begin{figure}[t!]
\centering
\includegraphics[width=0.5\textwidth]{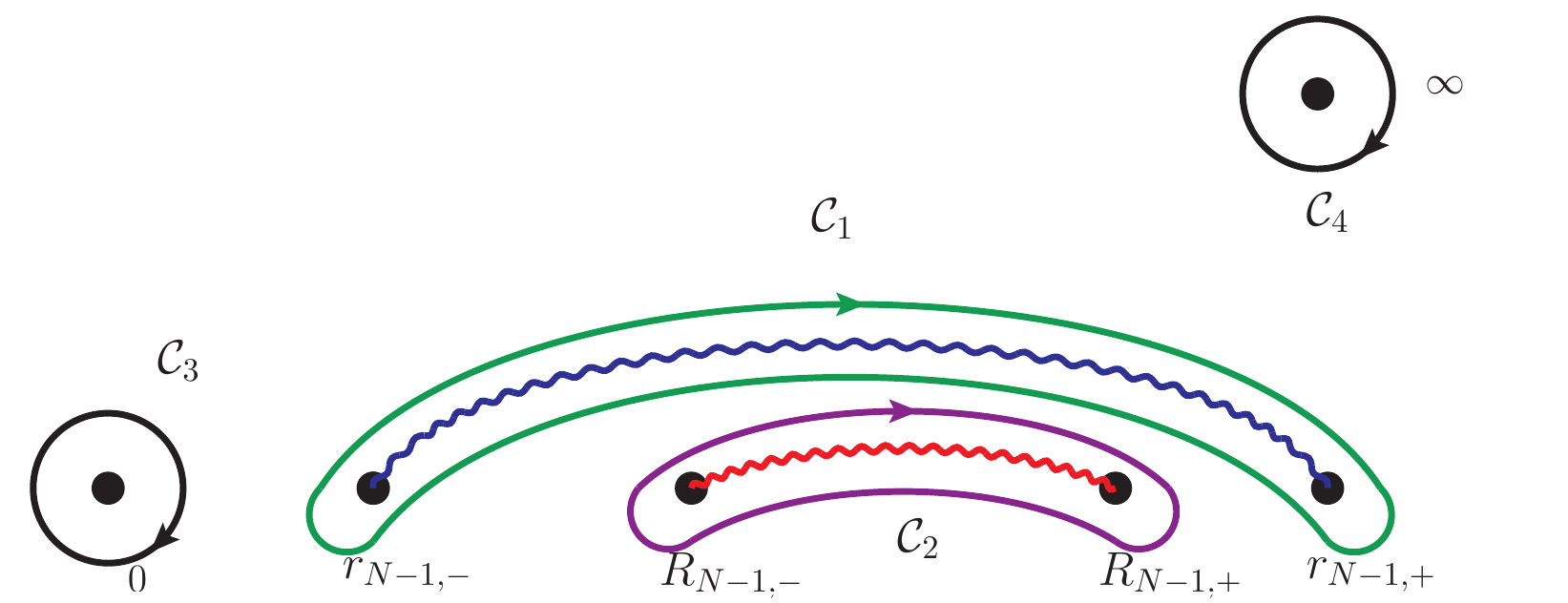}
\caption{\label{fig:contour_M53}The branch cuts and integration paths for $M_{N,N-2}$ with odd $N$.}
\end{figure}

We are now concerned with the singularities of the integrand $I(z_{N-1})$ in Eq.~\eqref{eq:I53}. There are two poles at $0$ and $\infty$, respectively. There is a branch cut between $r_{N-1,-}$ and $r_{N-1,+}$ for the square root. There is also a branch cut between $R_{N-1,-}$ and $R_{N-1,+}$ for the logarithm, where $R_{N-1,\pm}$ are the two roots of the polynomial $G_N(\bm{0}',z_{N-1},0)$. These singularities are depicted in Fig.~\ref{fig:contour_M53}. We define the integral path of Eq.~\eqref{eq:I53} to be the upper half of the contour $\mathcal{C}_1$. Hence we have
\begin{align}
I_{N,N-2} &= \frac{1}{8\pi} \int_{\mathcal{C}_1} I(z_{N-1}) \, dz_{N-1} \nn \\
&= -\frac{1}{8\pi} \int_{\mathcal{C}_2+\mathcal{C}_3+\mathcal{C}_4} I(z_{N-1}) \, dz_{N-1} \,.
\end{align}
The integration around $\mathcal{C}_3$ is just $(-2\pi i)$ multiplying the residue at $z_{N-1} = 0$, i.e.
\begin{align}
-\frac{1}{8\pi} \, d\int_{\mathcal{C}_3} I(z_{N-1}) \, dz_{N-1} &= \frac{i}{4} \, d\log\frac{\widetilde{B}_N - \sqrt{\mK_N \widetilde{G}_{N-1}^{(1)}}}{\widetilde{B}_N + \sqrt{\mK_N \widetilde{G}_{N-1}^{(1)}}}  \nn \\
&= \frac{i}{2} dM_{N,N-1}^{(1)} \,.
\end{align}
On the two sides of $\mathcal{C}_2$, the logarithm differs by $2\pi i$, and
\begin{align}
&-\frac{1}{8\pi} \, d\int_{\mathcal{C}_2} I(z_{N-1}) \, dz_{N-1} = \frac{i}{4} \, d\log\frac{R_{N-1,+}}{R_{N-1,-}} \nn \\
&= \frac{i}{4} \, d\log\frac{\widetilde{B}_N - \sqrt{\mK_N \widetilde{G}_{N-1}^{(2)}}}{\widetilde{B}_N + \sqrt{\mK_N \widetilde{G}_{N-1}^{(2)}}}  = \frac{i}{2} dM_{N,N-1}^{(2)} \,.
\end{align}

From the above we see that the genuine contribution to $dM_{N,N-2}$ only comes from the integration along $\mathcal{C}_4$. For that we need to investigate the behavior of the logarithm in Eq.~\eqref{eq:I53} in the limit $z_{N-1} \to \infty$. We first note that $G_{N-1}^{(1)}(\bm{0}',z_{N-1}) \sim -\mK_{N-2} z_{N-1}^2 / 4$ in that limit. As for $B_N(\bm{0}',z_{N-1},0)$, it is a linear function of $z_{N-1}$ and the coefficient can be extracted as
\begin{align}
&\frac{\partial B_N(\bm{0}',z_{N-1},0)}{\partial z_{N-1}} \nn \\
&= \frac{\partial G(l,p_1,\ldots,p_{E-1};\, p_E,p_1,\ldots,p_{E-1})}{\partial z_{N-1}} \nn
\\
&= \frac{\partial l \cdot p_E}{\partial z_{N-1}}\frac{\partial G(l,p_1,\ldots,p_{E-1};\, p_E,p_1,\ldots,p_{E-1})}{\partial l \cdot p_E} \nn
\\
&\quad + \frac{\partial l \cdot p_{E-1}}{\partial z_{N-1}}\frac{\partial G(l,p_1,\ldots,p_{E-1};\, p_E,p_1,\ldots,p_{E-1})}{\partial l\cdot p_{E-1}} \nn
\\
&= \frac{1}{2} G(p_1,\ldots,p_{E-1};\, p_1,\ldots,p_{E-1}) \nn
\\
&\quad + \frac{1}{2} G(p_1,\ldots,p_{E-2},p_{E-1};\, p_1,\ldots,p_{E-2},p_{E}) \nn
\\
&= \frac{1}{2}G(p_1,\ldots,p_{E-2},p_{E-1};\, p_1,\ldots,p_{E-2},p_{E-1}+p_E) \, .
\end{align}
We hence have
\begin{align}
dM_{N,N-2} &= -\frac{1}{8\pi} \, d\int_{\mathcal{C}_4} I(z_{N-1}) \, dz_{N-1} \nn \\
&= \frac{i}{4} d\log \frac{C_N - \sqrt{-\mK_N \mK_{N-2}}}{C_N + \sqrt{-\mK_N \mK_{N-2}}} \, ,
\label{eq:M53}
\end{align}
where
\begin{equation}
C_N = G(p_1,\ldots,p_{E-2},p_{E-1};\, p_1,\ldots,p_{E-2},p_{E-1}+p_E) \, .
\end{equation}
The letter $W_{N,N-2}$ can be readily read off. Note that the Gram determinants in this letter only involve external momenta. The letter hence has a well-defined limit when $\widetilde{G}_{N-2} = 0$ and $g_{N-2}$ is not a master. We will see what this means later.

\subsubsection{Even number of propagators}
\label{sec:W42}

If $N$ is an even number, assuming no divergence, we have the differential equation
\begin{align}
d\int_\mathcal{C} \frac{dz_{N-1}}{z_{N-1}} \frac{dz_N}{z_N}\frac{\sqrt{\widetilde{G}_N}}{\sqrt{G_N(\bm{0}',z_{N-1},z_N)}} = 4\pi \, dM_{N,N-2} \,,
\end{align}
where the domain $\mathcal{C}$ is determined by $G_N(\bm{0}',z_{N-1},z_N) \geq 0$. Note that the dependence on $g_{N-1}^{(i)}$ drops out in this case. We choose to integrate over $z_N$ first, and have
\begin{align}
dM_{N,N-2} &=\frac{1}{4\pi} \, d\int_{r_{N-1,-}}^{r_{N-1,+}}  \frac{dz_{N-1}}{z_{N-1}} \frac{\sqrt{\widetilde{G}_N}}{\sqrt{G_N(\bm{0}',z_{N-1},0)}} \nn \\
&\quad \times \int_{r_{N,-}}^{r_{N,+}} \frac{dz_N}{z_N}\frac{\sqrt{G_N(\bm{0}',z_{N-1},0)}}{\sqrt{G_N(\bm{0}',z_{N-1},z_N)}} \,,
\end{align}
where $r_{N,\pm}$ are the two roots of the polynomial $G_N(\bm{0}',z_{N-1},z_N)$ with respect to $z_N$ (treating $z_{N-1}$ as a constant). The integration range of $z_{N-1}$ is in turn determined by the discriminant $\Delta$ of $G_N(\bm{0}',z_{N-1},z_N)$ (with respect to the variable $z_N$). Writing $\Delta = \mK_4 G_{N-1}^{(1)}(\bm{0}',z_{N-1})$, we know the the bounds $r_{N-1,\pm}$ are just the two roots of the polynomial $G_{N-1}^{(1)}(\bm{0}',z_{N-1})$. Here we define
\begin{align}
G_{N-1}^{(1)}(\bm{z}',z_{N-1}) &= G(l,p_1,\ldots,p_{E-1}) \, , \nn \\
G_{N-1}^{(2)}(\bm{z}',z_{N}) &= G(l,p_1,\ldots,p_{E-1}+p_E) \, .
\end{align}

The integration over $z_N$ can be carried out using Eq.~\eqref{eq:pi}. We then arrive at
\begin{align}
dM_{N,N-2} &= \frac{i}{4} \, dI_{N,N-2} \,,
\end{align}
where
\begin{align}
I_{N,N-2} &= \int_{r_{N-1,-}}^{r_{N-1,+}}  \frac{dz_{N-1}}{z_{N-1}} \frac{\sqrt{\widetilde{G}_N}}{\sqrt{G_N(\bm{0}',z_{N-1},0)}} \, ,
\end{align}
where $r_{N-1,\pm}$ are the two roots of $G_{N-1}^{(1)}(\bm{0}',z_{N-1})$. We denote the two roots of $G_N(\bm{0}',z_{N-1},0)$ as $R_{N-1,\pm}$. We can then write
\begin{align}
&G_N(\bm{0}',z_{N-1},0) \nn \\
&= -\frac{1}{4} \mK_{N-1}^{(2)} (z_{N-1}-R_{N-1,+})(z_{N-1}-R_{N-1,-}) \, ,
\end{align}
where
\begin{equation}
\mK_{N-1}^{(2)} = G(p_1,\ldots,p_{E-2},p_{E-1}+p_E) \, .
\end{equation}
We define
\begin{equation}
t = \frac{1}{z_{N-1}} \, , \quad t_{\pm} = \frac{1}{r_{N-1,\mp}} \, , \quad T_{\pm} = \frac{1}{R_{N-1,\mp}} \, .
\end{equation}
The integral can then be written as
\begin{align}
I_{N,N-2} &= \int_{t_-}^{t_+} \frac{dt}{\sqrt{(t-T_+)(t-T_-)}} \nn \\
&= 2 \log \frac{\sqrt{t_+-T_+} + \sqrt{t_+-T_-}}{\sqrt{t_--T_+} + \sqrt{t_--T_-}} \, .
\end{align}

We now want to rewrite the above expression in terms of Gram determinants. To do that we first write
\begin{align}
G_N(\bm{0}',z_{N-1},0) &= -\frac{1}{4} \mK_{N-1}^{(2)} z_{N-1}^2 - \widetilde{B}_{N} z_{N-1} + \widetilde{G}_N \, , \nn
\\
G_{N-1}^{(1)}(\bm{0}',z_{N-1}) &= -\frac{1}{4}\mK_{N-2} z_{N-1}^2 - \widetilde{B}_{N-1}^{(1)} z_{N-1} + \widetilde{G}_{N-1}^{(1)} \, ,
\end{align}
where
\begin{align}
\mK_{N-2} &= G(p_1,\ldots,p_{E-2}) \, , \nn
\\
B_{N-1}^{(1)}(\bm{z}) &= G(l,p_1,\ldots,p_{E-2};\, p_{E-1},p_1,\ldots,p_{E-2}) \, .
\end{align}
The roots are given by
\begin{align}
t_{\pm} &= \frac{\widetilde{B}_{N-1}^{(1)} \pm \sqrt{\mK_{N-1}^{(1)} \widetilde{G}_{N-2}}}{2\widetilde{G}_{N-1}^{(1)}} \, , \nn
\\
T_{\pm} &= \frac{\widetilde{B}_{N} \pm \sqrt{\mK_{N} \widetilde{G}_{N-1}^{(2)}}}{2\widetilde{G}_{N}} \, ,
\end{align}
where
\begin{equation}
G_{N-1}^{(2)} = G(l,p_1,\ldots,p_{E-2},p_{E-1}+p_E) \, ,
\end{equation}
and we have used the relations
\begin{align}
B_N^2 + \mK_{N-1}^{(2)} G_N &= \mK_N G_{N-1}^{(2)} \, , \nn \\
\left( B_{N-1}^{(1)} \right)^2 + \mK_N G_{N-1}^{(1)} &= \mK_{N-1}^{(1)} G_{N-2} \, .
\end{align}

We can now employ the geometric representations of the Gram determinants in Eq.~\eqref{eq:triangle1} to simplify the expressions. Let $l^\star$ be the solution to $\bm{z}=0$, we will be concerned with the components of $l^\star$, $p_{E-1}$ and $p_{E-1}+p_E$ orthogonal to the subspace spanned by $\{p_1,\ldots,p_{E-2}\}$. For convenience we denote these components as $k^\mu$ (for $l^\star$), $p^\mu$ (for $p_{E-1}$) and $q^\mu$ (for $p_{E-1}+p_E$). We note that $k^\mu$ is time-like while $p^\mu$ and $q^\mu$ are space-like. Hence we can define the norms $|k| = \sqrt{k^2}$, $|p| = \sqrt{-p^2}$ and $|q| = \sqrt{-q^2}$. We further denote the components of $k^\mu$ and $p^\mu$ perpendicular to $q$ as $k_\perp^\mu$ and $p_\perp^\mu$, and define the corresponding norms as $|k_\perp|$ and $|p_\perp|$. We can finally write
\begin{align}
t_{\pm} = \frac{\sinh(\eta_1) \pm i}{2 |k| |p| \cosh^2(\eta_1)} \, , \quad T_{\pm} = \frac{\sinh(\eta_2) \pm i}{2 |k_{\perp}| |p_\perp| \cosh^2(\eta_2)} \, , 
\end{align}
where $\eta_1$ is the hyperbolic angle between $k$ and $p$, and $\eta_2$ is the hyperbolic angle between $k_\perp$ and $p_\perp$. It will be convenient to define the imaginary angle $\theta_{kp} \equiv \pi/2 - i\eta_1$, such that $\cosh(\eta_1) = \sin\theta_{kp}$ and $i \sinh(\eta_1) = \cos\theta_{kp}$; and similarly $\theta_{kp,\perp q} \equiv \pi/2 - i\eta_2$. 

We use $\theta_{pq}$ to denote the angle between $p$ and $q$, and define $\xi$ as the hyperbolic angle between $k$ and $q$ (with the corresponding imaginary angle $\theta_{kq} \equiv \pi/2 - i\xi$). We then have the relations
\begin{align}
|p_\perp| &= |p| \sin\theta_{pq} \, , \quad |k_\perp| = |k| \sin\theta_{kq} \, , \nonumber
\\
\cos\theta_{kp} &= \cos\theta_{kq} \cos\theta_{pq} + \cos\theta_{kp,\perp q} \sin\theta_{kq} \sin\theta_{pq} \, .
\label{eq:tri2}
\end{align}
It then follows that
\begin{align}
t_{\pm} - T_{\pm} &\equiv \frac{P_{\pm\pm}}{2|k_\perp| |p_\perp| \sin^2\theta_{kp} \sin^2\theta_{kp,\perp q}} \, ,
\end{align}
where
\begin{align}
P_{\pm\pm} &= (-i\cos\theta_{kp} \pm i) \sin^2\theta_{kp,\perp q} \sin\theta_{pq} \sin\theta_{kq} \nn \\
&\quad - (-i\cos\theta_{kp,\perp q} \pm i) \sin^2\theta_{kp} \, .
\end{align}
Plugging in the relation \eqref{eq:tri2}, we may write the functions $P_{\pm\pm}$ as
\begin{align}
P_{++} &= -8 i \, \sin^2\left( \frac{\theta_{kp}}{2} \right) \cos^2\left( \frac{\theta_{kq}+\theta_{pq}}{2} \right) \sin^2\left( \frac{\theta_{kp,\perp q}}{2} \right) , \nonumber
\\
P_{+-} &= 8 i \, \sin^2\left( \frac{\theta_{kp}}{2} \right) \cos^2\left( \frac{\theta_{kq}-\theta_{pq}}{2} \right) \cos^2\left( \frac{\theta_{kp,\perp q}}{2} \right) , \nonumber
\\
P_{-+} &= -8 i \, \cos^2\left( \frac{\theta_{kp}}{2} \right)  \sin^2\left( \frac{\theta_{kq}+\theta_{pq}}{2} \right) \sin^2\left( \frac{\theta_{kp,\perp q}}{2} \right) , \nonumber
\\
P_{--} &= 8 i \, \cos^2\left( \frac{\theta_{kp}}{2} \right) \sin^2\left( \frac{\theta_{kq}-\theta_{pq}}{2} \right) \cos^2\left( \frac{\theta_{kp,\perp q}}{2} \right) .
\end{align}
Using trigonometry identities together with the relations
\begin{align}
\cos\theta_{pq} &= \cos\theta_{kp} \cos\theta_{kq} + \cos\theta_{pq,\perp k} \sin\theta_{kp} \sin\theta_{kq} \, , \nonumber
\\
\sin\theta_{pq} &= \sin\theta_{pq,\perp k} \, \frac{\sin\theta_{kp}}{\sin\theta_{kp,\perp q}} \, ,
\end{align}
we can arrive at a surprisingly simple result
\begin{equation}
I_{N,N-2} = 2 \log e^{-i \theta_{pq,\perp k}} = \log \frac{\cos\theta_{pq,\perp k} - i \sin\theta_{pq,\perp k}}{\cos\theta_{pq,\perp k} + i \sin\theta_{pq,\perp k}} \, ,
\end{equation}
where $\theta_{pq,\perp k}$ is the angle between $p_{\perp k}$ and $q_{\perp k}$. It is straightforward to rewrite the above expression in terms of Gram determinants, and we finally obtain
\begin{align}
dM_{N,N-2} = \frac{i}{4} d\log \frac{\widetilde{D}_N - \sqrt{-\widetilde{G}_N\widetilde{G}_{N-2}}}{\widetilde{D}_N + \sqrt{-\widetilde{G}_N\widetilde{G}_{N-2}}} \,,
\label{eq:W42}
\end{align}
where $\widetilde{D}_{N} = D_N(\bm{0})$ and
\begin{equation}
D_N(\bm{z}) = G(l,p_1,\ldots,p_{E-1};l,p_1,\ldots,p_{E-1}+p_E) \, .
\end{equation}

\subsection{Dependence on further lower sub-sectors}

In the convergent case, $dg_{N}$ cannot depend on $g_{N-3}$ or integrals with even fewer propagators. For odd $N$, this can be easily seen from the powers of $\epsilon$ in Eq.~\eqref{eq:UTint}. For even $N$, however, $dg_{N}$ and $g_{N-3}$ are multiplied by the same power of $\epsilon$ in the differential equations. We then need to examine the three-fold integrals appearing in the differential equations under the $(N-3)$-cut. The first two folds can be performed following the calculations in Section~\ref{sec:W42}, and the last fold can be studied similar to Section~\ref{sec:W53}. Finally we can arrive at the conclusion that $dM_{N,N-3}=0$ in the convergent case. Note however, such dependence can be present in the divergence cases to be discussed in the next Section.

\section{Letters in differential equations: divergent cases}

We now consider the situation when some cut integrals become divergent and one cannot perform a Taylor expansion for the integrands. As discussed earlier, this happens when certain Gram determinants vanish under maximal cut, and the corresponding integrals are reducible to lower sectors. A classical example is the massless 3-point integral that can be reduced to 2-point integrals. Reducible higher-point integrals can occur with specific configurations of external momenta, which appear, e.g., at boundaries of differential equations or in some effective field theories. Divergent cut integrals can have two kinds of consequences, which we will discuss in the following.

\subsection{$N,N-2$ dependence with a reducible $(N-1)$-point integral}

We consider the dependence of $dg_{N}$ on $g_{N-2}$ when $g_{N-1}^{(1)}$ is reducible, where $N$ is even. Following the derivation in Section~\ref{sec:W42}, we see that now one of $r_{N-1,\pm}$ is zero and $G_{N-1}^{(1)}(\bm{0}',0) = 0$. The integration over $z_{N-1}$ is hence divergent and one cannot Taylor expand the integrand in $\epsilon$. One can also find that the entry $dM_{N,N-2}$ obtained in Section~\ref{sec:W42} is divergent. To proceed, we can keep the regulator in the differential equation:
\begin{align}
&d\int_\mathcal{C} \frac{dz_{N-1}}{z_{N-1}} \frac{dz_N}{z_N} \nn
\\
&\quad \times \frac{\sqrt{\widetilde{G}_N}}{\sqrt{G_N(\bm{0}',z_{N-1},z_N)}} \left[-\frac{\mK_N}{G_N(\bm{0}',z_{N-1},z_N)}\right]^\epsilon \nonumber
\\
&= \epsilon \, dM_{N} \int_\mathcal{C} \frac{dz_{N-1}}{z_{N-1}} \frac{dz_N}{z_N}  \nn \\
&\quad \times \frac{\sqrt{\widetilde{G}_N}}{\sqrt{G_N(\bm{0}',z_{N-1},z_N)}} \left[-\frac{\mK_N}{G_N(\bm{0}',z_{N-1},z_N)}\right]^\epsilon \nonumber
\\
&+ 4\pi \, dM_{N,N-2}^\star \left(-\frac{\mK_{N-2}}{\widetilde{G}_{N-2}}\right)^\epsilon + \mathcal{O}(\epsilon) \,,
\end{align}
where $dM_{N,N-2}^\star$ denotes the entry in the divergent case. Note that $g_{N-1}^{(1)}$ is not a master integral and does not contribute to the right-hand side, while the last $\mathcal{O}(\epsilon)$ denotes a suppressed contribution from another $(N-1)$-point integral $g_{N-1}^{(2)}$. Here we assume that $G_{N-1}^{(2)}(\bm{0}',0)$ is non-zero and the integration over $z_N$ is convergent for $\epsilon \to 0$.

We now need to perform Laurent expansions of the integrands in terms of distributions. We write
\begin{align}
G_{N-1}^{(1)}(\bm{0}',z_{N-1}) &= \frac{1}{4}\mK_{N-2} z_{N-1} \, (t - z_{N-1}) \, , \nn
\\
t &= - \frac{4\widetilde{B}_{N-1}^{(1)}}{\mK_{N-2}} \, .
\end{align}
We can then use
\begin{equation}
\int_0^t \frac{dz}{z^{1+\epsilon}} \, f(z) = -\frac{t^{-\epsilon}}{\epsilon} \, f(0) + \int_0^t \frac{dz}{z^{1+\epsilon}}  \left[ f(z) - f(0) \right] ,
\end{equation}
to perform the series expansion. In particular we have
\begin{align}
&\int_\mathcal{C} \frac{dz_{N-1}}{z_{N-1}} \frac{dz_N}{z_N} \nn
\\
&\quad \times \frac{\sqrt{\widetilde{G}_N}}{\sqrt{G_N(\bm{0}',z_{N-1},z_N)}} \left[-\frac{\mK_N}{G_N(\bm{0}',z_{N-1},z_N)}\right]^\epsilon \nonumber
\\
&= i\pi \int_0^t \frac{dz_{N-1}}{z_{N-1}^{1+\epsilon}} \frac{\sqrt{\widetilde{G}_N}}{\sqrt{G_N(\bm{0}',z_{N-1},0)}} \nn
\\
&\quad \times \left[ 1 + \epsilon \, h(z_{N-1}) + \mathcal{O}(\epsilon^2) \right] \nonumber
\\
&= i\pi \Bigg[ -\frac{1}{\epsilon} + \log(t) - h(0) \nn \\
&\quad + \int_0^t \frac{dz_{N-1}}{z_{N-1}} \left( \frac{\sqrt{\widetilde{G}_N}}{\sqrt{G_N(\bm{0}',z_{N-1},0)}} - 1 \right) \Bigg] + \mathcal{O}(\epsilon) \, ,
\label{eq:M42_div}
\end{align}
where the function $h(z_{N-1})$ arises from the expansion in $\epsilon$ after integrating over $z_N$. When $z_{N-1} \to 0$, it reduces to
\begin{equation}
h(0) = \log \left( \frac{4\mK_{N-1}^{(1)}}{\widetilde{B}_{N-1}^{(1)}} \right) + 4\log(2) \, .
\end{equation}

The last integral in Eq.~\eqref{eq:M42_div} can be obtained by taking the limit $\widetilde{G}_{N-1}^{(1)} \to 0$ in the difference between Eq.~\eqref{eq:W42} and a simple integral of $1/z_{N-1}$:
\begin{align}
&\int_0^t \frac{dz_{N-1}}{z_{N-1}} \left( \frac{\sqrt{\widetilde{G}_N}}{\sqrt{G_N(\bm{0}',z_{N-1},0)}} - 1 \right) \nn\\
&= \lim_{\widetilde{G}_{N-1}^{(1)} \to 0} \Bigg( \log\frac{\widetilde{D}_{N} -  \sqrt{-\widetilde{G}_{N} \widetilde{G}_{N-2}}}{\widetilde{D}_{N} +  \sqrt{-\widetilde{G}_{N} \widetilde{G}_{N-2}}} \nn \\
&\hspace{4em} - \log\frac{\widetilde{B}_{N-1}^{(1)} +\sqrt{\widetilde{G}_{N-2} \mK_{N-1}^{(1)}}}{\widetilde{B}_{N-1}^{(1)} -\sqrt{\widetilde{G}_{N-2} \mK_{N-1}^{(1)}}} \Bigg) \,.
\end{align}
Using the relations
\begin{gather}
-G_{N} G_{N-2}=D_{N}^2-G_{N-1}^{(1)}G_{N-1}^{(2)} \, , \nn \\
G_{N-2} \mK_{N-1}^{(1)}=\left(B_{N-1}^{(1)}\right)^2+G_{N-1}^{(1)}\mK_{N-2} \, ,
\label{eq:W42_div_G1}
\end{gather}
we can simplify the expression and arrive at
\begin{multline}
\int_0^t \frac{dz_{N-1}}{z_{N-1}} \left( \frac{\sqrt{\widetilde{G}_N}}{\sqrt{G_N(\bm{0}',z_{N-1},0)}} - 1 \right) 
\\
= \log \frac{\widetilde{G}_N  \mK_{N-2}}{\mK_{N-1}^{(1)}\widetilde{G}_{N-1}^{(2)} } \, .
\end{multline}
Now we can combine everything and find in the divergent case (for even $N$) that
\begin{equation}
W_{N,N-2}^\star = \frac{\widetilde{G}_{N-2} \, \mK_N}{\mK_{N-1}^{(1)} \, \widetilde{G}_{N-1}^{(2)}} \, .
\label{eq:W42}
\end{equation}

Comparing to Eq.~\eqref{eq:W42}, we note that the letter in the divergent case is simpler (without square roots) than that in the convergent case. Interestingly, this simple letter can be obtained without going through the tedious calculation in the above. We observe that in the divergent case $\widetilde{G}_{N-1}^{(1)} \to 0$, we have the relation
\begin{equation}
\tilde{g}_{N-1}^{(1)} = -\frac{1}{2} \tilde{g}_{N-2} \, .
\end{equation}
This hints that we should combine $dM_{N,N-1}^{(1)}$ and $dM_{N,N-2}$ to arrive at $dM_{N,N-2}^\star$:
\begin{align}
dM_{N,N-2}^{\star} &= \lim_{\widetilde{G}_{N-1}^{(1)}\to 0} \left( -\frac{1}{2} dM_{N,N-1}^{(1)} + dM_{N,N-2} \right) \nn
\\
&= \frac{i}{4} \lim_{\widetilde{G}_{N-1}^{(1)} \to 0} \Bigg( \log\frac{\widetilde{D}_{N} -  \sqrt{-\widetilde{G}_{N} \widetilde{G}_{N-2}}}{\widetilde{D}_{N} +  \sqrt{-\widetilde{G}_{N} \widetilde{G}_{N-2}}} \nn
\\
&\hspace{4em} - \log\frac{\widetilde{B}_{N}^{(1)} -\sqrt{-\widetilde{G}_{N} \mK_{N-1}^{(1)}}}{\widetilde{B}_{N}^{(1)} +\sqrt{-\widetilde{G}_{N} \mK_{3}^{(N)}}} \Bigg) \,.
\end{align}
Using the relations in Eq.~\eqref{eq:W42_div_G1} as well as
\begin{align}
-G_{N} \mK_{N-1}^{(1)}&=\left(B_{N}^{(1)}\right)^2+G_{N-1}^{(1)}\mK_{N} \, ,
\label{eq:gramdetforM42IR}
\end{align}
we can easily arrive at Eq.~\eqref{eq:W42}.

Further divergences may arise if $\widetilde{G}_{N-1}^{(2)} = 0$ in Eq.~\eqref{eq:W42}. In this case both $g_{N-1}^{(1)}$ and $g_{N-1}^{(2)}$ are reducible to lower-point integrals. It can be shown that the corresponding letter can be obtained by including $dM_{N,N-1}^{(2)}$, and we do not elaborate on the calculation here. We finally note that the above considerations can also be applied to the $N$-odd cases, although here $g_{N-1}^{(i)}$ can only be reducible for specific configurations of external momenta. We will encounter similar situations in the next subsection.

\subsection{$N,N-3$ dependence with a reducible $(N-2)$-point integral}

In the convergent case, we have seen that $dg_{N}$ can only depends on $g_N$, $g_{N-1}^{(i)}$ and $g_{N-2}^{(i)}$. This picture changes in the divergent case when one of $g_{N-2}^{(i)}$ is reducible, and $dg_{N}$ may develop dependence on some $(N-3)$-point integrals. As a practical example, we consider the dependence of 5-point integrals on 2-point ones. According to Eq.~\eqref{eq:DE-letter}, we have
\begin{align}
d\tilde{g}_5 &= \epsilon \, dM_5 \, \tilde{g}_5 + \epsilon \sum_i dM_{5,4}^{(i)} \, \tilde{g}_4^{(i)} + \epsilon \sum_i dM_{5,3}^{(i)} \, \tilde{g}_3^{(i)} \nn
\\
&+ \epsilon \, dM_{5,2} \, \tilde{g}_2 \, ,
\end{align}
where the cut on $z_1$ and $z_2$ is imposed. Using Eq.~\eqref{eq:UTint} we arrive at
\begin{align}
dM_{5,2} + \mathcal{O}(\epsilon) &= \frac{\epsilon}{8\pi} \, dI_{5,2}(\epsilon) - \frac{\epsilon}{4\pi} \sum_{i=3}^5 dM_{5,4}^{(i)} \, I_{4,2}^{(i)}(\epsilon) \nn \\
&- \frac{\epsilon}{2} \sum_{i=3}^5 dM_{5,3}^{(i)} \, I_{3,2}^{(i)}(\epsilon) \, ,
\label{eq:M52eq}
\end{align}
where
\begin{align}
I_{5,2}(\epsilon) &= \int \frac{dz_3}{z_3} \frac{dz_4}{z_4} \frac{dz_5}{z_5} \left( -\frac{\mK_5}{G_5(0,0,z_3,z_4,z_5)} \right)^\epsilon , \nonumber
\\
I_{4,2}^{(i)}(\epsilon) &= \int \frac{dz_j}{z_j} \frac{dz_k}{z_k} \frac{\sqrt{G_4^{(i)}(0,0,0,0)}}{\sqrt{G_4^{(i)}(0,0,z_j,z_k)}} \nn
\\
&\hspace{4em} \times \left( -\frac{\mK_4^{(i)}}{G_4^{(i)}(0,0,z_j,z_k)} \right)^\epsilon , \nonumber
\\
I_{3,2}^{(i)}(\epsilon) &= \int \frac{dz_i}{z_i} \left( -\frac{\mK_3^{(i)}}{G_3^{(i)}(0,0,z_i)} \right)^\epsilon ,
\end{align}
where $j<k$ and $j,k \neq i$. We note that there is a factor of $\epsilon$ in each term on the right-hand side of Eq.~\eqref{eq:M52eq}. Therefore the term can only contribute if the integral is divergent in the limit $\epsilon \to 0$. For that to happen, at least one of $G_3^{(i)}(0,0,0)$ needs to vanish. For simplicity we assume $G_3^{(3)}(0,0,0) = 0$, while the other two $G_3^{(i)}(0,0,0)$'s are non-zero. In any case, it is clear that the $I_{3,2}^{(i)}(\epsilon)$ terms do not contribute, since they are either zero or non-divergent. The integrals $I_{4,2}^{(4)}(\epsilon)$ and $I_{4,2}^{(5)}(\epsilon)$ are similar to Eq.~\eqref{eq:M42_div} with the result $-i\pi/\epsilon + \mathcal{O}(\epsilon^0)$. Therefore we only need to deal with the divergent part of $I_{5,2}(\epsilon)$:
\begin{align}
I_{5,2}(\epsilon) &= \int \frac{dz_3}{z_3} \frac{dz_4}{z_4} \left[ \Delta(z_3,z_4) \right]^{-\epsilon} \nn \\
&\log\frac{B_5^{(5)}(0,0,z_3,z_4,0) - \sqrt{\Delta(z_3,z_4)}}{B_5^{(5)}(0,0,z_3,z_4,0) + \sqrt{\Delta(z_3,z_4)}} + \mathcal{O}(\epsilon^0) \, ,
\end{align}
where
\begin{equation}
\Delta(z_3,z_4) = \mK_5 G_{4}^{(5)}(0,0,z_3,z_4) \, .
\end{equation}
The integration over $z_4$ is similar to Eq.~\eqref{eq:I53}, except the additional factor $\Delta^{-\epsilon}$, which regularizes the divergence as $z_3 \to 0$. Since we are only interested in the leading term in $\epsilon$, it is equivalent to replace this factor by $z_3^{-\epsilon}$. We can then expand $z_3^{-1-\epsilon}$ in terms of distributions. Keeping only the $1/\epsilon$ terms, we have
\begin{align}
&dI_{5,2}(\epsilon) + \mathcal{O}(\epsilon^0) \nn \\
&= -\frac{1}{\epsilon} d\int \frac{dz_4}{z_4} \log\frac{B_5^{(5)}(0,0,0,z_4,0) - \sqrt{\Delta(0,z_4)}}{B_5^{(5)}(0,0,0,z_4,0) + \sqrt{\Delta(0,z_4)}} \nn\\
\nn \\
&= -\frac{1}{\epsilon}\left( 2\pi i \,  dM_{5,4}^{(4)} + 2\pi i\, dM_{5,4}^{(5)} + 4\pi \, dM_{5,3}^{(3)} \right)\, ,
\end{align}
where the second line follows from the calculation of Eq.~\eqref{eq:I53}. We finally arrive at
\begin{equation}
dM_{5,2} = -\frac{1}{2} dM_{5,3}^{(3)} = -\frac{i}{8} d\log \frac{C_5 - \sqrt{-\mK_5 \mK_{3}}}{C_5 + \sqrt{-\mK_5 \mK_{3}}}
\label{eq:W52}
\end{equation}
where
\begin{equation}
C_5 = G(p_1,p_2,p_3,p_4;\, p_1,p_2,p_3,p_4+p_5) \, .
\end{equation}

The result in Eq.~\eqref{eq:W52} is unsurprising given the relation $g_3^{(3)}=-g_2/2$. Similar behaviors are observed when more than one $\widetilde{G}_3$ vanish. The corresponding $dM_{5,2}$ is then a linear combination of several $dM_{5,3}$'s.
We hence conclude that letters in these cases can also be obtained straightforwardly without tedious calculations.

The above discussion relates the appearance of $dM_{N,N-3}$ to the reducibility of one or more $g_{N-2}^{(i)}$'s. One may imagine that, if in addition, one or more $g_{N-3}^{(i)}$'s becomes reducible, there can be $dM_{N,N-4}$ appearing in the differential equations. This is impossible for integrals with generic external momenta (i.e., the $E$ external momenta are indeed independent). However, such cases may arise at certain boundaries of kinematic configurations. When this happens, the corresponding letters can be easily obtained following the reduction rules among the integrals, as was done in the previous paragraph.

\section{Summary and outlook}
\label{sec:summary}

In summary, we have studied the alphabet for one-loop Feynman integrals. The alphabet governs the form of the canonical differential equations, and provides important information on the analytic solution of these equations. We find that the letters in the alphabet can be generically constructed utilizing the UT integrals in the Baikov representation under various cuts. We first considered cases where all the cut integrals are convergent in the limit $\epsilon \to 0$. The corresponding letters coincide with the results in \cite{Abreu:2017ptx, Abreu:2017enx, Abreu:2017mtm}, while our expressions are simpler in certain cases. We have also thoroughly studied the cases of divergent cut integrals. We find that letters in the divergent cases can be easily obtained from the convergent cases by taking certain limits. The letters admit universal expressions in terms of various Gram determinants. We have checked our general results in several known examples, and found agreements. We have also applied our results to the complicated case of a $2 \to 3$ amplitude with 7 physical scales. The details about that is presented in Ref.~\cite{Chen:2022nxt}.

We expect that our results will be useful in many calculations of $2 \to 3$ and $2 \to 4$ amplitudes which are theoretically and/or phenomenologically interesting. It is also interesting to see whether similar universal structures can be obtained at higher loop orders, using the UT integrals in the Baikov representation of \cite{Chen:2020uyk, Chen:2022lzr}.

\begin{acknowledgments}
This work was supported in part by the National Natural Science Foundation of China under Grant No. 11975030, 11635001 and 11925506.
\end{acknowledgments}

\bibliographystyle{apsrev4-1}
\bibliography{references_inspire.bib}

\end{document}